# Experimental and theoretical analysis of noise strength and environmental correlation time for ensembles of nitrogen-vacancy centers in diamond

Kan Hayashi[1*], Yuichiro Matsuzaki[2], Takaki Ashida[1], Shinobu Onoda[3], Hiroshi Abe[3], Takeshi Ohshima[3], Mutsuko Hatano[4], Takashi Taniguchi[5], Hiroki Morishita[1], Masanori Fujiwara[1], Norikazu Mizuochi[1]

[1]Instutute for Chemical Research, Kyoto University, Uji, Kyoto 611-0011, Japan
[2]Nanoelectronics Research Institute, National Institute of Advanced Industrial Science and Technology (AIST), 1-1-1 Umezono, Tsukuba, Ibaraki 305-8568, Japan.
[3]National Institutes for Quantum and Radiological Science and Technology (QST), Takasaki, Gunma 370-1292, Japan
[4]Department of Physical Electronics, Tokyo Institute of Technology, 2-12-1 Ookayama, Meguro-ku, Tokyo 152-8552, Japan
[5]National Institute for Materials Science (NIMS), Tsukuba, Ibaraki 305-0044, Japan

Abstract

NV centers in diamond has been considered as an important building block for the realization of quantum information processing such as quantum simulation, quantum memory, and quantum metrology. To maximize the potential of the NV centers, it is essential to understand the mechanism of the decoherence. Existing theories predicted a relationship between the coherence time and spin concentration (a sum of the concentration of the NV center and P1 center) in diamond. However, a systematic experimental study of the spin concentration dependence of the coherence time was still missing. Here, we experimentally and theoretically investigate the Hahn echo decay curve with several diamond samples with different spin concentration. The Hahn echo results show that we observe a non-exponential decay for the low spin concentration while an exponential decay is dominant for the high spin concentration. By fitting the decay curve with a theoretical model, we show that both the amplitude and correlation time of the environmental noise has a clear dependence on the spin concentration. These results are essential to optimize the NV center concentration as high-performance quantum devices, particularly as quantum sensors.

## 1. Introduction

Nitrogen vacancy centers (NV centers) in diamond are a promising system for realizing quantum information processing [1-10,17–36]. An NV center is a spin 1 system. Magnetic fields parallel to the NV center axis lift the degeneracy between $m_s$ = +1 and $m_s$ = -1 by a Zeeman shift $\Delta E = \gamma B$, where $\gamma$ is the gyromagnetic ratio for the electron spin (28 MHz/mT) and $B$ is the magnetic field parallel to the NV center axis. Microwaves with a frequency of around 2.87 GHz can induce a transition between $m_s$ = 0 and $m_s$ = ±1 for the electronic ground state. By using frequency selectivity, we can use this system as an effective two-level system. The electronic ground state for the NV center spin ($S$ = 1) can be initialized and read by optical pumping, allowing us to perform optically detected magnetic resonance (ODMR) with NV centers [1]. Moreover, NV centers have a long coherence time and stability even at room temperature [2,3]. These properties indicate NV centers have the potential to realize quantum information processing involving quantum memory [4–7], quantum simulation [8–10], and distributed quantum computation [11–16]. One of the most attractive applications of NV centers is for high sensitivity magnetometers [17–29]. A single NV center allows nanoscale magnetometry with a reasonably high sensitivity (~ nT/Hz$^{1/2}$) [2,3,17,18], making it possible to perform nano-NMR [19]. For a sensor with a size of about a micron, where magnetoencephalography and magnetocardiogram applications are possible, an ensemble of NV centers has a high sensitivity proportional to the square root of the number of NV centers [20–24]. A sensitivity of 0.9 pT/Hz$^{1/2}$ has been demonstrated with 1.4 × 10$^{11}$ NV centers using spin echo based magnetometry [20]. Also, several techniques such as the use of a sophisticated pulse sequence [26–29] or hybridization with other systems [4–7] have expanded the potential of NV based magnetometers. Furthermore, theoretical calculations have shown that the optimal sensitivity is around 250 aT/$\sqrt{\text{Hz}}$/cm$^{2/3}$, which is comparable to that of SQUIDs [17], though the sensitivity of NV magnetometers is currently lower than the optimal value and still needs to be improved.

Importantly, NV centers suffer from decoherence, which is an obstacle for more practical applications. By applying a π/2-pulse, we can create a superposition between $m_s$ = 0 and $m_s$ = ±1. However, the coherent superposition of two states accumulates a relative phase of the target magnetic field, and low frequency fluctuations of the magnetic field induce dephasing of the NV centers. Importantly, a spin echo pulse sequence eliminates such environmental noise and thus improves the coherence time [2,3,17,18]. This technique is widely used in quantum information processing with NV centers [2,3,17,18]. With a high-density ensemble of NV centers, dipole interactions from nitrogen paramagnetic impurities (P1 centers) are a dominant noise source for spin dephasing [17,18,30–34]. Interactions between the NV centers are another strong noise source [17,34]. To realize a practical sensor, it is essential to understand how the nitrogen concentration affects decoherence, which will let us determine a strategy to improve the coherence time. There are many ways to suppress decoherence, such as the quantum Zeno effect [35–39], dynamical decoupling [30,45,46], qubit motion [47,48], and quantum error correction [40–43]. However, the efficiency of these protocols strongly depends on the noise properties. For example, the quantum Zeno effect and qubit motion can be used when quadratic decay is experimentally observed, while these schemes do not improve the coherence time for exponentially decaying systems. Dynamical decoupling can be used to suppress decoherence only when the environment is non-Markovian. Quantum error correction can detect noise and recover quantum states for any type of local decoherence [48,49]. However, when using NV centers for quantum sensors, quantum error correction could suppress not only decoherence but also signal accumulation from the target fields, and so a careful assessment of the environment and choice of error correction code is necessary in order to improve the sensitivity when using quantum error correction [40–43].

Also, when using NV centers for entanglement enhanced magnetic field sensors, it is crucial to understand the origin of decoherence [50,51]. Recently, a systematic way to create a few NV centers within a distance of tens of nanometers has been developed [52], which paves the way to realizing a sensitive quantum sensor with entangled NV centers. However, from a theoretical point of view, it is known that an entanglement sensor is more effective than a classical sensor (that uses separable states) only when the decoherence behavior shows non-exponential decay [50,51]. This means that for practically useful entanglement sensors, understanding and controlling the environment is necessary.

In this article, we report a systematic study of Hahn echo decay curves for NV centers in several diamond samples with different spin concentrations. Our experimental results show that an exponential decay curve is observed for a high nitrogen concentration and a non-exponential decay curve is observed for a low nitrogen concentration. Using our model, we can fit the decay curve for both low and high NV center concentrations. Our analysis also shows that the correlation time and amplitude of the dipole interaction from the environmental spin bath have a linear dependence on the spin concentration.

## 2. Experiment

Nine sample diamond crystals with substitutional nitrogen (P1) centers were used in our study. The diamonds were synthesized by high-pressure high-temperature (HPHT) processes [53]. Among them, samples No. 5–9 were commercially obtained. The NV centers were created through 2-MeV electron irradiation at various doses (the irradiation doses are shown in Table 1) at 745±10 °C and subsequent annealing at 1000 °C for 1 hour in Ar gas. After annealing, the diamonds were cleaned in a mixture of boiling acids (1:1 sulfuric, nitric) to remove graphitic carbon and oxygen-termination at the surface. The P1 center concentrations were evaluated by electron paramagnetic resonance (EPR) spectroscopy at room temperature after electron irradiation and annealing. The NV center concentrations were investigated by comparing the fluorescent intensity between a single NV center and our samples. The fluorescence intensity was measured by applying a 532-nm laser with a lab-built confocal microscope (detection volume: 0.04 µm$^3$). Table 1 shows both the P1 and NV concentrations for the nine diamond samples, estimated from the EPR and fluorescence measurements. The Hahn echo sequence (Fig. 1(a)) for the diamonds was performed at room temperature. Green laser pulses at 50 µW illuminated the diamond samples. The laser pulse length is 100 µsec (No.1 ~No.4) and 50 µsec (No.5 ~No.9). Microwave pulses were applied by a copper wire above the diamond surface. The NV centers had a Rabi frequency $\Omega = 2\pi \times 8.3$ MHz for a $\pi$ pulse duration around 60 nsec. The magnetic field (~30 mT) was applied along the <111> direction using a permanent magnet to separate the revival peaks from the Hahn echo decay.

## 3. Result and discussion

We fit our experimental results using an analytical expression for the decay curves. First, we fit the results with a simple stretched exponential function ($\exp(-(\frac{t}{T_2})^n)$) and estimated the coherence time ($T_2$). Fig. 1(b) shows an almost linear dependence on the spin concentration (the sum of the NV center concentration and P1 center concentration). As the environment to decohere the NV centers, there are two types of the spin bath: resonant spins (such as NV centers with the same axis) and non-resonant spins (such as NV centers with different axis and P1 centers). With the high density of the NV center, $T_2$ is limited by not only dipole interactions from non-resonant spins but also instantaneous diffusion ($T_{ID}$), which is

the interaction between resonant spins after the microwave pulse, given by [54]

$$\frac{1}{T_{ID}} = n \frac{\pi}{9\sqrt{3}} \frac{\mu_0 g_1 g_2 \beta_e^2}{\hbar} \sin^2 \frac{\beta}{2} \tag{1}$$

Where $n$ is the density of the resonant spins, $g_1$ and $g_2$ are the $g$ factor of the resonant spins, $\hbar$ is the Dirac's constant, $\mu_0$ is the permeability of vacuum, $\beta_e$ is the Bohr magneton, and $\beta$ is the flip angle of the microwave pulse. In our case, $n$ is equal to the 1/12 of the density of the NV center due to the hyperfine coupling of the $^{14}$N nuclear spins and the applied magnetic fields along to a specific NV axis. From this consideration, we can estimate the decoherence rate affected by the non-resonant spin ($T_{non-reso}$) by using $\frac{1}{T_{non-reso}} = \frac{1}{T_2} - \frac{1}{T_{ID}}$. Fig. 1(c) show the linear dependence of the $1/T_{non-reso}$ on the non-resonant spin concentration (the difference between the spin concentration and $n$). Although the stretched exponential function is useful for estimating $T_2$ of the NV center, this fit does not capture some important properties of the environment. So, we adopted a random classical Gaussian noise model that contains the noise amplitude and correlation time of the environment, as we will describe later. This model is used to study the noise properties of the spin bath of a single NV center [30]. We introduce the decoherence of a single qubit when we perform a spin echo [17]. The interaction Hamiltonian between the qubit and the environment is considered to be

$$H_I(t) = \lambda f'(t) \hat{\sigma}_z \tag{2}$$

$$f'(t) = f(t) g(t) \tag{3}$$

where $g(t)$ is a modulation function defined as $g(t) = 1$ for $t \leq \tau/2$ and $g(t) = -1$ for $t > \tau/2$, $\lambda$ is the amplitude of the noise and $f(t)$ is a classical Gaussian random variable. Here, $\lambda$ is related to $T_2^*$ for the NV centers. If the source of the noise during the free induction decay is only the electron spin bath, we have the relationship $T_2^* = 1/2\pi\lambda$.

Also, we adopt a correlation function $\overline{f(t)f(0)} = e^{-\frac{1}{\tau_c}|t|}$, where $\tau_c$ is the correlation time of the environmental spin bath. This correlation function has been used to describe the spin bath for an NV center in diamond [30]. We consider non-biased noise such that $\overline{f(t)} = 0$. From this model, we can derive the Hahn echo decay curve as (see the supplementary information for details):

$$\langle 1|\overline{\rho_I(\tau)}|0\rangle = \frac{1}{2}e^{-4\lambda^2\left(\tau_c\tau + \left(-3 - e^{-\frac{\tau}{\tau_c}} + 4e^{-\frac{\tau}{2\tau_c}}\right)(\tau_c)^2\right)} \qquad (4)$$

For $\tau \ll \tau_c$, we have $\langle 1|\overline{\rho_I(\tau)}|0\rangle \simeq \frac{1}{2}e^{-\frac{\lambda^2 t^3}{3\tau_c}}$, while we have $\langle 1|\overline{\rho_I(\tau)}|0\rangle \simeq \frac{1}{2}e^{-4\lambda^2\tau_c\tau}$ for $\tau \gg \tau_c$. These results are consistent with previous work [11]. Here, we assumed the non-resonant spin and resonant spins has the same dependence on $\lambda$ and $\tau_c$ for the simplified analysis. In Fig. 3, we show the fitting results for the Hahn echo decay curves. We observed periodic revival peaks due to rephasing of the $^{13}$C spin bath [55,56]. Since we are interested in the decoherence effect from an electron spin bath that induces a monotonic decay, we performed a fit of the envelope of the data. The theoretical calculations show that we can observe a non-exponential decay for a long correlation time while an exponential decay should be dominant for a short correlation time. Actually, our experiments show a non-exponential decay with lower density samples while the decay becomes exponential for higher density samples. Intuitively, this shows that if the density of the electron spin is increased, the correlation time should be shorter. For a more quantitative analysis, we plot the values of $\tau_c$ and $\lambda$, obtained from fitting our experimental data using Eq. (4) or (11), with respect to the spin concentration (the sum of the NV center concentration and P1 center concentration) in Fig. 3. These parameters show an almost linear dependence on the spin concentration. Note that this is the first systematic and quantitative investigation of how the Hahn echo decay curve depends on the spin concentration.

Let us discuss a possible contribution of a flip-flop interaction between the NV centers for the decoherence. To observe the effect of the flip-flop interaction between the NV centers, the coupling between the nearest NV centers should be comparable with the detuning between them. However, it would be difficult to satisfy such a condition in our setup for the following reasons. First, due to the hyperfine coupling of the $^{14}$N nuclear spins, the frequency of the NV centers is split up into three frequencies whose separation is around 2.3 MHz. Second, due to the applied magnetic fields, only 25% of the NV centers are involved in the dynamics and the other NV centers are well detuned. Third, the P1 center induces an inhomogeneous broadening that provides the detuning between the NV centers. Actually, by considering these three conditions, we can perform an order estimation of the effect of the dipole-dipole interaction as follows. The coupling strength between the NV centers can be estimated from the density of the NV centers. For example, with a density of $10^{18}$ /cm$^3$, the dipole-dipole interaction is around tens of kHz. However, due to the first and second reasons described above, the effective coupling strength for the flip-flop interaction between the nearest NV centers should be 12 times smaller than what is estimated from the actual density. On the other hand, the inhomogeneous broadening due to the P1 center can be also estimated from the similar calculations above where the dipole-dipole interaction is around tens of kHz with a density of $10^{18}$ /cm$^3$ as an example. Therefore, as long as the density of the P1 center is comparable with or larger than that of the NV centers, the inhomogeneous broadening effect from the P1 center is at least one order of magnitude larger than the flip-flop interaction, and we can ignore the flip-flop interaction, which is the case of our setup.

Finally, we discuss possible future work. Regarding fabrication, there is room to improve the coherence time for the ensemble of NV centers. Especially, our HPHT samples contain some residual metal impurities that decrease the coherence time for the NV centers. Moreover, by decreasing the $^{13}$C nuclear spins, we could eliminate unwanted oscillations in the spin echo signal. For our theoretical model, where we have adopted a semi-classical noise model, we could instead use a microscopic model that can partially include many-body quantum effects [56]. While our semi-classical model can successfully reveal a clear relation between the noise parameters and spin concentration, a fully quantized model could give a deeper understanding of the decoherence mechanism for the NV centers.

## 4. Conclusion

We have investigated the Hahn echo decay curve for nine samples of ensembles of NV centers with different spin concentrations. Our results show a crossover from an exponential decay for high density samples to a non-exponential decay for low density samples. Our theoretical analysis of these experimental data reveals that the crossover between the forms of the decay is due to a change in the noise amplitude and environmental correlation time. By fitting the data with a theoretical decay curve, we show that the noise amplitude and the environmental correlation time have a linear dependence on the spin concentration of the diamond sample. Our results are crucial for understanding the decoherence dynamics of NV centers in diamond with various spin densities and will be useful for developing applications of NV centers for quantum information processing.

## 5. ACKNOWLEDGMENTS

This work was supported by the Leading Initiative for Excellent Young Researchers MEXT Japan, and in part supported by MEXT KAKENHI (Grant Nos. 15H05868, 15H05870, 16H06326), CREST (JPMJCR1333) and MEXT Q-LEAP. *Note added.* -While we are preparing our manuscript, we become aware of a related work that also shows a relationship between the spin density and coherence time with NV centers by Bauch et al [57].

## 6. Appemdix

We solve the Schrodinger equation in an interaction picture to obtain

$$\rho_I(\tau) = \rho(0) - i \int_0^\tau dt_1 [H_I(t_1), \rho(t_1)]$$

$$= \rho(0) + \sum_{n=1}^{\infty} (-i)^n \int_0^\tau \int_0^{t_1} \cdots \int_0^{t_{n-1}} dt_1 dt_2 \cdots dt_n \left[ H_I(t_1), \left[ H_I(t_2), \cdots, [H_I(t_n), \rho(0)] \right] \right] \quad (5)$$

By taking the average, we obtain

$$\overline{\rho_I(\tau)} = \rho(0) + \sum_{n=1}^{\infty} (-i\lambda)^n \int_0^\tau \int_0^{t_1} \cdots \int_0^{t_{n-1}} dt_1 dt_2 \cdots dt_n \overline{f'(t_1) \cdots f'(t_n)} \left[ \hat{\sigma}_z, \left[ \hat{\sigma}_z, \cdots, [\hat{\sigma}_z, \rho(0)] \right] \right]$$

$$= \rho(0) + \sum_{n=1}^{\infty} \frac{(-i\lambda)^n}{n!} \int_0^\tau \int_0^\tau \cdots \int_0^\tau dt_1 dt_2 \cdots dt_n \overline{f(t_1) \cdots f(t_n)} g(t_1) \cdots g(t_n) \left[ \hat{\sigma}_z, \left[ \hat{\sigma}_z, \cdots, [\hat{\sigma}_z, \rho(0)] \right] \right]$$

For a random Gaussian variable, all the cumulants except the second-order terms tend to be zero. Thus, we can ignore all moments of odd order, giving

$$\overline{f'(t_1) \cdots f'(t_{2m})} = (2m-1)(2m-3) \cdots 3 \cdot 1 \left( \overline{f'(t_1) f'(t_2)} \right)^m \quad (7)$$

So, we obtain

$$\overline{\rho_I(\tau)} = \rho(0) + \sum_{m=1}^{\infty} \frac{(-i\lambda)^{2m}}{m!} \frac{1}{2^m} \int_0^\tau \int_0^\tau dt_1 dt_2 \left( \overline{f(t_1)f(t_2)} g(t_1) g(t_2) \right)^m \left[ \hat{\sigma}_z, \left[ \hat{\sigma}_z, \cdots, [\hat{\sigma}_z, \rho(0)] \right] \right]_{2m} \quad (8)$$

Here, we assume $\overline{f(t_1)f(t_2)} = e^{-\frac{t_1-t_2}{\tau_c}}$, and we obtain

$$\frac{1}{2}\int_0^\tau \int_0^\tau dt_1 dt_2 (\overline{f(t_1)f(t_2)} g(t_1)g(t_2))$$
$$= \int_0^\tau \int_0^{t_1} dt_1 dt_2 (e^{-\frac{t_1-t_2}{\tau_c}} g(t_1)g(t_2))$$
$$= \tau_c \tau + \left(-3 - e^{-\frac{\tau}{\tau_c}} + 4e^{-\frac{\tau}{2\tau_c}}\right)(\tau_c)^2 \qquad (9)$$

Finally, we obtain

$$\overline{\rho_I(\tau)} = \rho(0) + \sum_{m=1}^\infty \frac{(-i\lambda)^{2m}}{m!}\left(\tau_c\tau + \left(-3 - e^{-\frac{\tau}{\tau_c}} + 4e^{-\frac{\tau}{2\tau_c}}\right)(\tau_c)^2\right)^m \left[\hat{\sigma}_z, \left[\hat{\sigma}_z, \cdots, [\hat{\sigma}_z, \rho(0)]\right]\right]_{2m} \qquad (10)$$

For an initial state $|+\rangle = \frac{1}{\sqrt{2}}(|0\rangle + |1\rangle)$, the non-diagonal terms of the density matrix show the decay curve, and we describe the Hahn echo decay curve as

$$\langle 1|\overline{\rho_I(\tau)}|0\rangle = \frac{1}{2}e^{-4\lambda^2\left(\tau_c\tau + \left(-3 - e^{-\frac{\tau}{\tau_c}} + 4e^{-\frac{\tau}{2\tau_c}}\right)(\tau_c)^2\right)} \qquad (11)$$

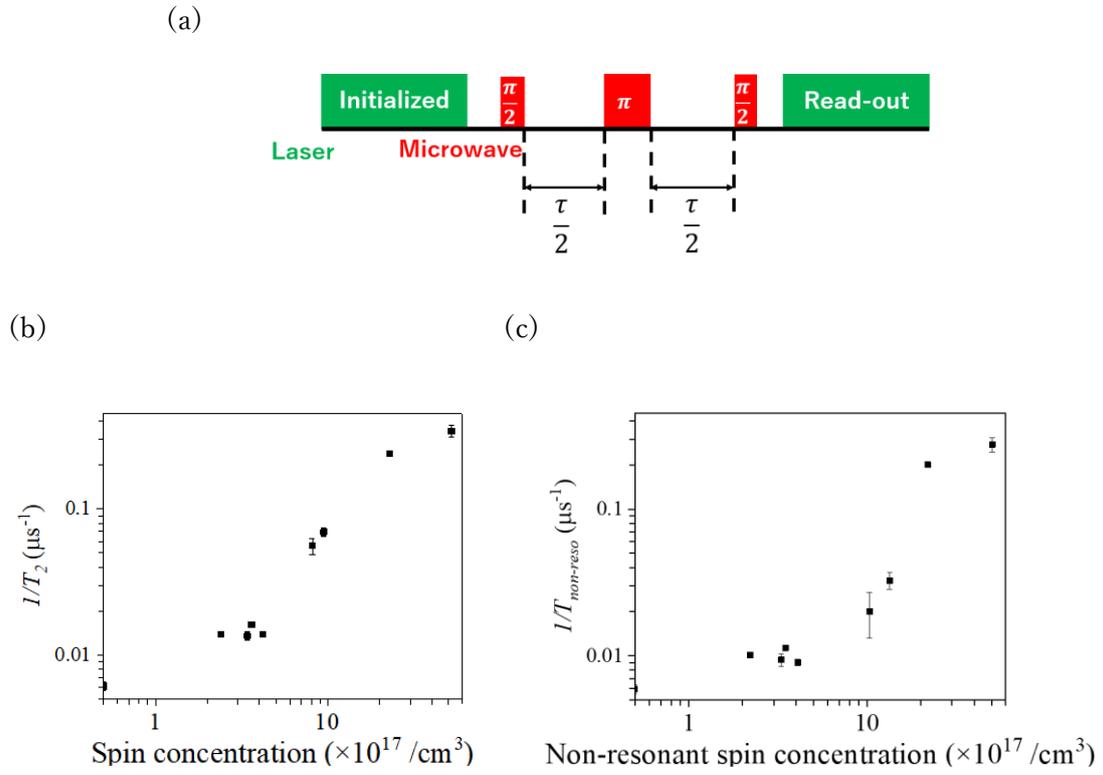

Fig. 1. (a) Typical Hahn echo sequence. (b) $T_2$ estimated by fitting experimental results with stretched exponential decay curve. (c) $T_{non-reso}$ estimated by $\frac{1}{T_{non-reso}} = \frac{1}{T_2} - \frac{1}{T_{ID}}$

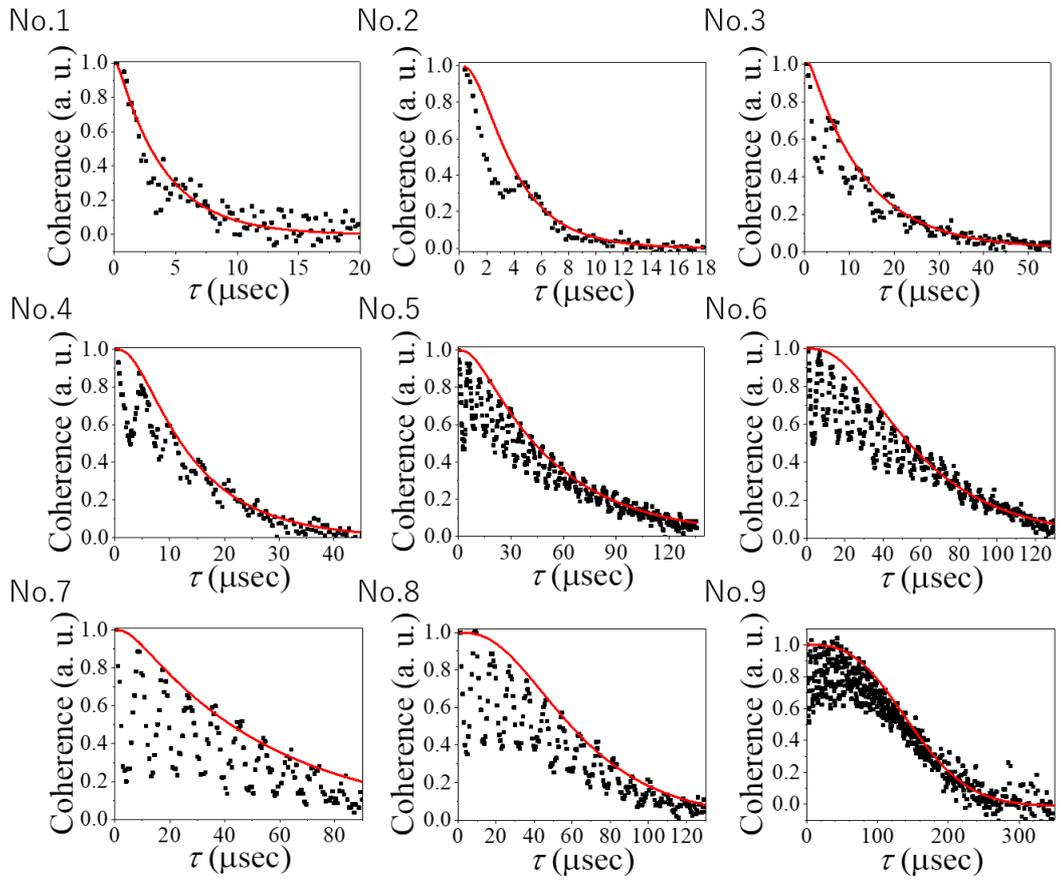

Fig. 2. Fitting results for Hahn echo decay curve for all diamond samples. The black and red lines are the experimental and numerical results, respectively. Eq. (10) was used for the fitting. Suitable magnetic fields were applied in the diamonds <111> direction.

(a)  (b)

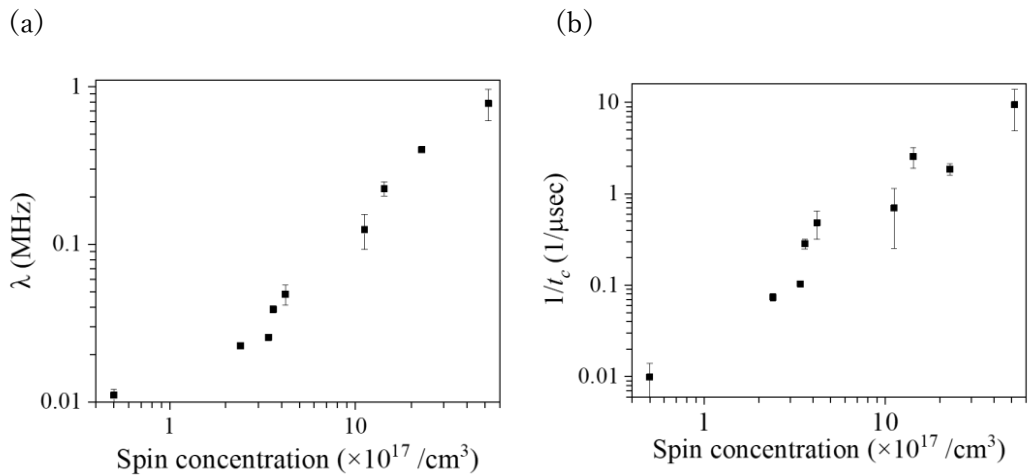

Fig. 3. Plots of (a) $\lambda$ and (b) $1/t_c$ with respect to spin concentration. The parameters were estimated by fitting the experimental results using Eq. (10).

Table 1

|  | P1 concentration ($\times 10^{17}$ /cm$^3$) | NV concentration ($\times 10^{17}$ /cm$^3$) | Electron irradiation dose ($\times 10^{16}$ e/cm$^2$) |
|---|---|---|---|
| No. 1 | 33.6 | 18.4 | 100 |
| No. 2 | 11.6 | 11.2 | 50 |
| No. 3 | 3.6 | 10.7 | 100 |
| No. 4 | 0.8 | 10.4 | 100 |
| No. 5 | 2.2 | 1.4 | 10 |
| No. 6 | 2.2 | 1.2 | 4 |
| No. 7 | 2.8 | 1.4 | 4 |
| No. 8 | 1.2 | 1.1 | 4 |
| No. 9 | 0.44 | 0.06 | 0.7 |

Table 1. P1 center concentration and NV center concentration of HPHT samples.